\documentstyle[aps,multicol,prb,epsf]{revtex} 
\begin{document} 
\draft
\title{The magnetic susceptibility of disordered non-diffusive
       mesoscopic systems} 
\author{Edward McCann\cite{lancaster} and Klaus Richter} 
\address{Max-Planck-Institut f\"{u}r Physik komplexer Systeme,
N\"othnitzer Str. 38, 01187~Dresden, Germany} 
\date{\today} 
\maketitle
\begin{abstract} 
{Disorder-induced spectral correlations of mesoscopic quantum systems 
in the non-diffusive regime and their effect on the magnetic 
susceptibility are studied.
 We perform impurity averaging for non-translational invariant systems 
by combining a diagrammatic perturbative approach with semiclassical 
techniques.
 This allows us to study the entire range from clean to diffusive systems.
As an application we consider the magnetic response of non-interacting
electrons in microstructures in the presence of weak disorder.
We show that in the ballistic case
 (elastic mean free path $\ell$ larger than the system size)
there exist two distinct regimes of behaviour 
depending on the relative magnitudes of $\ell$ and an inelastic 
scattering length $L_{\phi}$.
We present numerical results for square billiards and derive
approximate analytical results for generic chaotic geometries.
 The magnetic field dependence and $L_{\phi}$ dependence of 
the disorder-induced susceptibility is qualitatively similar 
in both types of geometry.}
\end{abstract} 
\pacs{03.65.Sq,05.45.+b,05.30.Fk,73.20.Dx}
%
\def \gplus#1{{\cal G}^+ \left( #1 \right) }
\def \gminus#1{{\cal G}^- \left( #1 \right) }
\def \diff#1{{\cal D} \left( #1 \right) }
\def \lp {L_{\phi}}
\def \xo {x_1}
\def \xop {x_1^{\prime}}
\def \xt {x_2}
\def \yo {y_1}
\def \yop {y_1^{\prime}}
\def \yt {y_2}

%
\bibliographystyle{simpl1}
\begin{multicols}{2}
%
%
\section{Introduction}
Mesoscopic physics has traditionally involved the study of novel phenomena in 
phase-coherent, disordered conductors \cite{AltLeeWebb:91}.
In the diffusive regime, where the elastic mean free path $\ell$ 
is much smaller than the system size $L$, the electron motion 
resembles that of a random walk between impurities.

The more recent development of high-mobility semiconductor heterostructures
together with advanced lithographic techniques has led to 
the confinement of electrons in two-dimensional 
microstructures of controllable, non-random geometry.
Thereby, a further regime, coined  ``ballistic'' since $\ell > L$ 
has been realised. Such technological achievements have motivated
theoretical approaches where the actual microcavities are approximated
by ``clean'' quantum billiards ignoring impurity scattering completely.
In these models the electron motion is only affected by bounces 
at the confinement potential, which is the opposite case to the 
diffusive regime where confinement effects are not important on 
timescales shorter than the Thouless time.

However residual impurity scattering is nearly unavoidable 
even in high-mobility microstructures. It has been shown that weak disorder 
can be strong enough to mix energy levels, influence spectral statistics 
\cite{S+I:86,A+G:93} and affect related thermodynamic quantities
\cite{Ric:96} in the ``ballistic'' regime.

In defining spectral correlation functions in the ``ballistic'' regime 
one has to distinguish between disorder 
averaging, which we denote by $\left< \ldots \right>_d$,
and size averaging, $\left< \ldots \right>_L$, which we assume to
be equivalent to energy averaging.
Pure disorder averaging corresponds to the experimental
situation of an ensemble of weakly disordered microstructures 
of the same size. Recent work has shown that for this case spectral 
correlation functions contain, and often are dominated by, strong 
oscillatory structures in energy (on scales $k_FL$, $k_F$ being the
Fermi wavenumber) reflecting the presence of the
confinement\cite{A+G:93,Ric:96}. They can be semiclassically 
interpreted as density-of-states oscillations related to classical 
periodic orbits of the corresponding clean system.

In the present work we consider spectral correlations after both
energy and (independent) disorder averaging corresponding to the experimental
situation of an ensemble of disordered systems with variation in their sizes.
Then one is able to divide the two-level correlation function, 
$K(\varepsilon_1,\varepsilon_2)$, 
into two separate terms, \cite{A+G:93,McC+Ri:98}
\begin{equation}
K(\varepsilon_1,\varepsilon_2 ;H) \equiv 
\left<\, K^d(\varepsilon_1,\varepsilon_2 ;H) \,\right>_L
 + K^L(\varepsilon_1,\varepsilon_2 ;H).
\label{ktot}
\end{equation}
Here 
\begin{mathletters}
\begin{eqnarray}
\label{kd}
K^d(\varepsilon_1,\varepsilon_2 )   & = &
\left[
\langle \nu(\varepsilon_1) \nu(\varepsilon_2) \rangle_d 
- \langle \nu(\varepsilon_1) \rangle_d 
\langle \nu(\varepsilon_2) \rangle_d 
\right] / \bar{\nu}^2
 \; , \\
K^L(\varepsilon_1,\varepsilon_2)  & = & 
\langle \langle \nu(\varepsilon_1) \rangle_d 
\langle \nu(\varepsilon_2) \rangle_d \rangle_L
/ \bar{\nu}^2 - 1 \; ,
\label{kl}
\end{eqnarray}
\end{mathletters}
where $\nu$ denotes the single particle density of states 
and $\bar{\nu} = \langle \langle \nu(\varepsilon) \rangle_d \rangle_L$
its mean part.  $K^d$ is a measure of 
{\em disorder-induced} correlations of the density of states, 
while $K^L$ is given by {\em size-induced} correlations.
In a diffusive system, $\ell < L$, the disorder-averaged 
density of states $\left<\nu (\varepsilon )\right>_d$ is a constant, 
so that $K^L$ is vanishingly small and $K^d$ dominates. 
However, for $\ell > L$ the density of states contains terms
which oscillate like $\cos (k_F L)$,
and both correlation functions may be relevant.
In contrast, once $\ell > L (k_FL)^{d-1}$ disorder-induced mixing of levels
is negligible and $K^L$ prevails.

The orbital magnetism of mesoscopic quantum systems is sensitive
to spectral correlations as e.g.\ measured by $K(\varepsilon_1,\varepsilon_2)$ 
and therefore has been the subject of much theoretical interest\cite{orbmag}
as well as experimental investigations\cite{percon,Mailly,Lev:93}.
For isolated systems with a fixed number of particles 
it is necessary to consider averaging under canonical conditions 
\cite{Che:88,B+M:89,Alt:91} resulting in a large
contribution to the average magnetic response which can be by orders of 
magnitude larger than the (bulk) Landau diamagnetism.
The corresponding susceptibility is given by \cite{Alt:91}
\begin{equation}
\label{def_sus}
\left<\chi (H)\right> = - \left(\frac{\Delta }{2} \right) 
\frac{\partial^2}{\partial H^2}
\left<\delta N^2 (\mu ;H)\right>
\end{equation}
for temperatures $k_{\rm B}T$ larger than the mean level spacing $\Delta$.
In Eq.~(\ref{def_sus}) $H$ is the magnetic field and 
$\left<\delta N^2 (\mu ;H)\right>$ is the variance in the
number of energy levels within an energy interval of width 
equal to the chemical potential, $\mu$.  This variance is related  to
$K(\varepsilon_1,\varepsilon_2 ;H)$, Eq.~(\ref{ktot}), by integration of the
level energies $\varepsilon_1$, $\varepsilon_2$ over the energy interval.
Throughout this work we will label the contributions to the susceptibility, 
corresponding to $\left<\, K^d \,\right>_L$ and $K^L$, as
$\left<\chi^d (H)\right>$ and $\left<\chi^L (H)\right>$, 
respectively. Hence the total orbital magnetic susceptibility is composed of
\begin{equation}
\left<\chi (H)\right> = \left<\chi^d (H)\right>
+ \left<\chi^L (H)\right>.
\end{equation}

Diagrammatic techniques to treat impurity scattering
are usually designed for diffusive systems,
$\ell <L$, and therefore  
rely on the assumption of translational invariance.
In this paper we combine a diagrammatic perturbation approach with
semiclassical techniques to perform energy and impurity averaging  
and to calculate the disorder induced part of
the energy correlation function and the related susceptibility 
$\left<\chi^d (H)\right>$ for disordered microstructures 
in the non-diffusive regime. In this quantum-semiclassical hybrid approach,
scattering at the impurity potentials, which are assumed to be $\delta$-like,
is treated quantum mechanically in a perturbation series with respect 
to the disorder.
Boundary effects are incorporated in a semiclassical representation of
the Green functions, which enter into the impurity diagrams,
in terms of classical paths. This method, therefore,
takes into account, in a systematic way, 
contributions from (closed) trajectories
which involve both scattering at impurities and specular reflection at the
confinement potential. 
The procedure allows us to study the complete crossover from diffusive  
to clean systems. A preliminary brief account of this work has been presented
in Ref.~\cite{McC+Ri:98}.

Our approach carries features of the ``method of trajectories'' originally
devised for thin superconducting films \cite{G+T:64}.
Note that similar \cite{B+H:88}, and alternative \cite{D+K:84}, 
methods have been also employed to consider weak localization in 
thin films in a parallel magnetic field. Related semiclassical approaches
for diffusive systems were proposed by Chakravarty and Schmid\cite{C+S:86}
and Argaman et al.\cite{AIS:93}.
More recently, Agam and Fishman\cite{Agam:96} studied 
the spectral form factor for ballistic systems with rigid disks (spheres) 
as impurities from a quantum chaos point of view.
We further note that a non-perturbative approach for ballistic
systems has been developed by Muzykantskii and Khmelnitskii,
the ballistic $\sigma$ model\cite{M+K:95}.

For real systems, besides $\ell$, there are additional relevant lengthscales 
at which inelastic scattering ($L_{\phi}$) or temperature smearing ($L_{T}$) 
produce a damping of propagation.
For clarity we refer to such a lengthscale in the following as $L_{\phi}$, 
although we assume that similar general arguments will hold for finite $L_{T}$. 
For ballistic motion at the Fermi energy $E_F$, we can relate $L_\phi$
to the level broadening $\gamma$ by 
\begin{equation}
\frac{L_\phi}{L} = \frac{k_F L}{2\pi} \frac{\Delta}{\gamma} 
\label{Lphi} \; .
\end{equation}
It  divides the ``ballistic'' regime into two sub-regimes.
In the first, $L , L_{\phi} < \ell$, the particle motion is 
nearly ballistic: the damping (of the Green function)
due to inelastic scattering typically occurs before elastic impurity 
scattering; for the remainder of this
paper we refer to this regime as {\it inelastic}.
 In the second regime, $L < \ell < L_{\phi}$, a particle may scatter many 
times off impurities before scattering inelastically. We call 
this regime {\it elastic}.  In our semiclassical treatment we consider
the contribution of orbits {\em of all lengths} 
(smaller than $v_F t_H$, where $t_H \approx \hbar /\Delta$ 
is the Heisenberg time)
in both the elastic and inelastic regime.

We present and compare results for non-interacting ballistic quantum 
systems with integrable and chaotic classical dynamics in the clean limit.
We derive analytical estimates for the average susceptibility 
$\left<\chi^d (H)\right>$ of generic chaotic microstructures
assuming ergodicity for the classical paths involved.
As an example of an integrable geometry, we treat in detail the 
representative case of the square billiard.
Experiments \cite{Lev:93} on the magnetic susceptibility
of ensembles of squares were performed in the ``ballistic'' 
inelastic regime, motivating theoretical studies of the susceptibility 
for $L < \ell$ \cite{Ric:96,Gef:94}.
Gefen, Braun, and Montambaux (GBM) \cite{Gef:94} considered 
the contribution of trajectories longer than $\ell$ to 
$\left<\chi^d (H)\right>$ in an approximate way, 
finding a paramagnetic $\ell$-independent contribution at zero field, 
whereas Richter, Ullmo, and Jalabert (RUJ) \cite{Ric:96} 
calculated $\left<\chi^L (H)\right>$ for a square by assuming that 
the disorder perturbs the phase, but not the trajectory, of 
semiclassical paths of the corresponding clean geometry. 
In the elastic regime we find agreement with the results 
of GBM with regard to the $k_F L$ behavior, while the zero field 
susceptibility is weakly dependent on $\ell$.
In the inelastic regime, for larger level broadening, we find 
entirely different results, namely an exponential dependence 
on $L/L_{\phi}$ and on $\ell /L$.

The present paper is organized as follows: In the next section
we outline our combined semiclassical diagrammatic technique. In
Sec.~\ref{sec:square} we illustrate our approach by applying it
 to the case of integrable (square) billiards.
We give analytical results for spectral correlation functions
(at zero magnetic field) and numerical results for the
magnetic susceptibility. 
In Sec.~\ref{sec:chaos} we present the derivation of the
corresponding disorder induced spectral correlations and 
susceptibility for chaotic ballistic geometries.
%
%
%
%
\section{Semiclassical diagrammatic approach}
\label{sec:sc_dia}
In this section we  summarize our semiclassical 
evaluation of the disorder correlation function $K^d$, Eq.~(\ref{kd}),
and the corresponding susceptibility $\left<\chi^d (H)\right>$.
To this end we begin with the diagrammatic formulation of the
problem and then perform the semiclassical approximations.
\subsection{Diagrammatic framework}
\label{subsec:diag}

We consider non-interacting electrons in a 
weak, perpendicular magnetic field and a random Gaussian 
potential $V$, with $\left<V({\bf r})\right>=0$ and correlator
\begin{equation}
\left<V({\bf r})\,V({\bf r^\prime})\right>=
\frac{\hbar}{2\pi\bar{\nu}\tau}\,\delta({\bf r}-{\bf r^\prime})\,,
 \label{wn}
\end{equation}
describing white noise disorder. In Eq.~(\ref{wn})
$\tau$ is the mean elastic scattering time, $\ell = v_F\tau$, 
and $\bar{\nu} = m L^2/2\pi\hbar^2$ in two dimensions.
In terms of the retarded and advanced single particle Green functions
obeying the boundary conditions of the corresponding clean system, 
${\cal G}^{+(-)}({\bf  r}_1 , {\bf  r}_2 ; \varepsilon ;H)$, the
correlator $K^d$ may be written as 
\begin{eqnarray}
K^d(\varepsilon_1,\varepsilon_2 ;H) &\approx&
\left( \frac{\Delta^2}{2\pi^2}\right) {\cal R} \left<\!\left<
{\rm tr} \,{\cal G}^{+}(\varepsilon_1 ;H) {\rm tr} \,
{\cal G}^{-}(\varepsilon_2 ;H)
\right>\!\right>_d \; . \nonumber \\
&& \label{discor}
\end{eqnarray}
Here, the average is taken over impurities only (for a given system size) and 
the symbol $\left<\!\left< \ldots \right>\!\right>_d$ implies 
the inclusion of connected diagrams only. 

\begin{figure}
\vspace{0.3cm}
\hspace{0.02\hsize} 
\epsfxsize=0.9\hsize
\epsffile{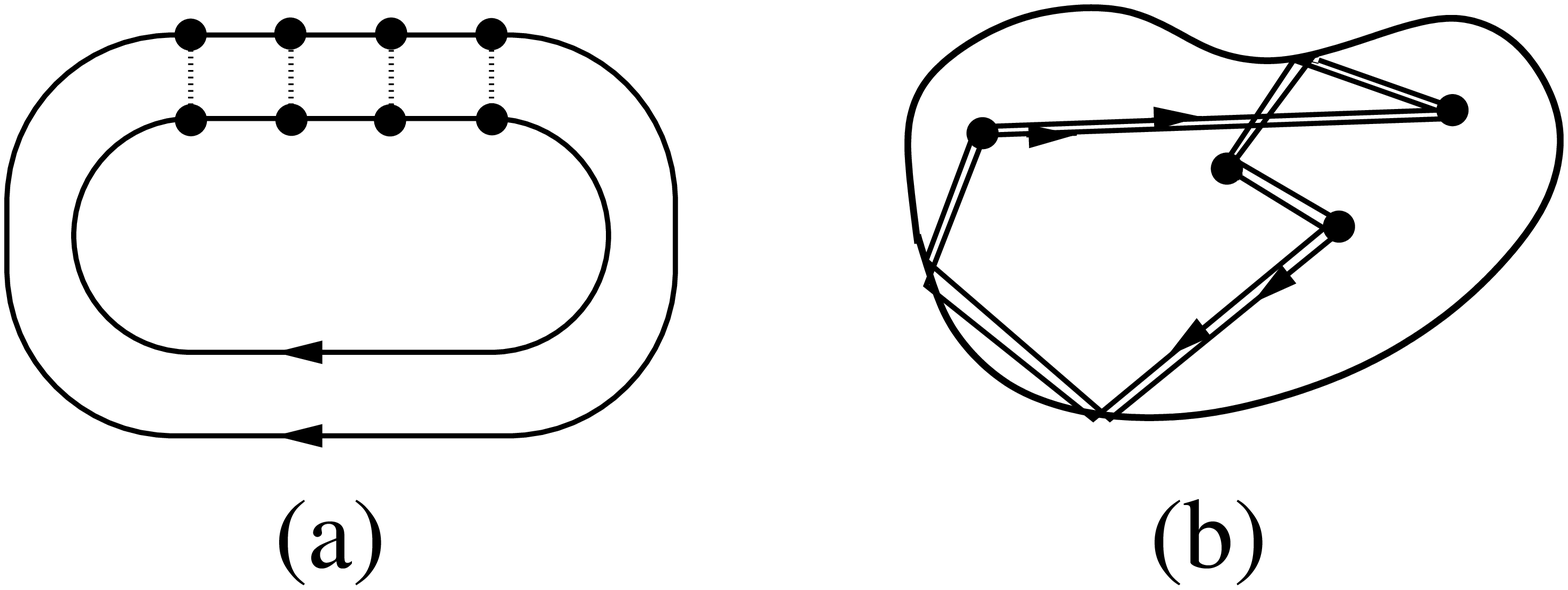}
\vspace{0.3cm}
 \refstepcounter{figure}
\label{DIAG}
{\small \setlength{\baselineskip}{10pt} FIG.\ \ref{DIAG}.
The diagrams ${\cal S}_{n}^{(C)}$,
 (a) shows a schematic form of the diagram ${\cal S}_{4}^{(C)}$, 
(b) shows a pair of typical real space semiclassical trajectories which 
contribute to ${\cal S}_{4}^{(C)}$.}
\end{figure}

We are particularly interested in the field sensitive part, 
$\tilde{K}^d$, of $K^d$.  Following a diagrammatic approach 
introduced in Ref.~\cite{A+G:93} it can be expressed as 
\begin{equation}
\label{Kdiag}
\tilde{K}^d (\epsilon_1, \epsilon_2; H) = \frac{\Delta^2}{2\pi^2}
\frac{\partial}{\partial \epsilon_1}\frac{\partial}{\partial \epsilon_2}
 {\cal R} \sum_{n=1}^\infty 
\frac{1}{n} {\cal S}_n^{(C)}(\omega; H) 
\end{equation}
with $\omega = \varepsilon_1 - \varepsilon_2$.
The Cooperon type diagrams ${\cal S}_{n}^{(C)}$ are defined by
\begin{eqnarray}
\label{sn}
{\cal S}_{n}^{(C)}(\omega ;H) & = & 
 {\rm Tr} \left[\zeta^{(C)}(\omega; H)\right]^n  \\
&=& \left(\int \prod_{j=1}^{n} d^dr_j\right) \prod_{m=1}^{n}
\zeta^{(C)} ({\bf r}_m,{\bf r}_{m+1};\omega ;H) \nonumber
\end{eqnarray}
with ${\bf r}_{n+1} \equiv {\bf r}_1$ and
\begin{eqnarray}
\lefteqn{
\zeta^{(C)} ({\bf  r}_1 , {\bf  r}_2 ;\omega ;H) =\qquad\qquad }  \nonumber \\
& & \frac{\hbar}{2\pi\bar{\nu}\tau}
G^{+}({\bf  r}_1 , {\bf  r}_2 ; \varepsilon_1;H ) \
G^{-}({\bf  r}_1 , {\bf  r}_2 ; \varepsilon_2;H ) \; .
\label{zc}
\end{eqnarray}
Here $G^{+(-)} = \left< {\cal G}^{+(-)} \right>_d$ is the disorder 
averaged single particle Green function.  An example, 
${\cal S}_{4}^{(C)}$, is shown schematically in Fig.~\ref{DIAG}a.
 Note that the sum of diagrams ${\cal S}_{n}^{(C)}$ is equivalent to a 
number of one-loop diagrams in the conventional notation: they include 
the dominant contribution to $\tilde{K}^d$ in the diffusive regime 
for $d > 2$ (Ref.~\cite{A+S:86}) 
but are actually smaller than some two-loop diagrams in a wide region 
of energies in the diffusive regime at $d=2$ (Ref.~\cite{K+L:95}).
However in the region of interest of this paper, 
the non-diffusive ``ballistic'' regime, 
the diagrams ${\cal S}_{n}^{(C)}$ dominate.
%
%
%
%
%
%
%
%
\subsection{Semiclassical treatment}
\label{subsec:sc}

Due to the lack of translational invariance in confined systems it is
no longer convenient to evaluate diagrams such as that in Fig.~\ref{DIAG}
in momentum space. Instead we work in configuration space and compute
the integrals (\ref{sn}) invoking a semiclassical approximation.

Semiclassically, the single particle Green function 
${\cal G}^+({\bf r}_1, {\bf r}_2)$ 
can be expressed as a sum over classical trajectories $t$ between 
${\bf  r}_1$ and ${\bf  r}_2$\cite{Gut:90}. After disorder average
it reads\cite{Ric:96}
\begin{equation}
 G^{+}({\bf  r}_1 , {\bf  r}_2) \simeq 
 \sum_{t:{{\bf  r}_1} \rightarrow {{\bf r}_2}} \
D_t \exp{\left(\frac{i}{\hbar} S_t({\bf r}_1,{\bf r}_2)-
             \frac{L_t}{2\ell}\right)} \; .
\label{scGreen}
\end{equation}
The classical amplitude $D_t$ includes the local density of trajectories
near the path $t$, $S_t$ stands for the classical action along the orbit 
including the Maslov index, and $L_t$ is the orbit length.

Eq.~(\ref{scGreen}) was derived in a semiclassical framework
under the assumption that weak disorder
modifies merely the phases $S_t$, which leads to damping on the scale of
$\ell$, leaving the trajectories unaffected; i.e.\ the sum is taken over
the paths of the corresponding clean system. 

For the bulk case this
treatment is equivalent to the usual diagrammatic treatment of 
disorder\cite{Ric:96}:
For white noise the self-consistent Born approximation leads to the
following integral equation for the impurity averaged
single particle Green function\cite{Stone}:
\begin{eqnarray}
 G^{+}({\bf  r}_1 , {\bf  r}_2) & = &  {\cal G}^+({\bf r}_1, {\bf r}_2) 
+  \\
 & + &   \frac{\hbar}{2\pi\bar{\nu}\tau} 
 \int  {\rm d}  {\bf r}_3 \ 
  {\cal G}^+({\bf r}_1, {\bf r}_3) G^{+}({\bf  r}_3 , {\bf  r}_3) 
 G^{+}({\bf  r}_3 , {\bf  r}_2) \; .  \nonumber
\end{eqnarray}
In the bulk case $G^{+}({\bf  r}_3 , {\bf  r}_3)$ in the above equation
is given in terms of ``paths of zero length'' leading to the 
exponential damping of the (free) Green function without disorder.
For confined systems further paths of finite length may semiclassically
contribute to $G^{+}({\bf  r}_3 , {\bf  r}_3)$.
They can be viewed as paths scattered off the impurities which may
lead to additional corrections to the impurity averaged
single particle Green function. 
However, these contributions are of higher order 
in $\hbar$ and $1/\tau$\cite{M+R}.

Upon using the semiclassical expression (\ref{scGreen}) in Eq.~(\ref{zc})
the two-particle operator
$\zeta^{(C)}({\bf  r}_1,{\bf  r}_2;\omega ;H)$ is then  given 
in terms of a double sum over pairs of classical paths which explicitly 
include the effect of boundary scattering.
However most pairs (of different paths) produce oscillating 
contributions which we assume to vanish after energy or size 
averaging \cite{note3}.
Therefore the main contribution to the field sensitive part of 
$\left<\, K^d \,\right>_L$ arises from diagonal terms 
(otherwise known as the Cooperon channel) obtained by pairing paths with 
their time reverse.  Assuming that the magnetic field affects 
the phase of the particles but not their trajectories 
we can write for the actions
\begin{eqnarray}
\lefteqn{
\frac{1}{\hbar} S_t(\epsilon_i; H) \simeq } \nonumber \\
& & \frac{1}{\hbar} S_t(E_F; H\!=\!0) 
+ (\epsilon_i -E_F) T_t +  \frac{2\pi}{\varphi_0}
\int_{{\bf  r}_1}^{{\bf  r}_2} {\bf A.dr}  \; ,
\label{S_expan}
\end{eqnarray}
where $T_t$ is the period of the trajectory, $\bf A$ is the vector potential
and $\varphi_0 = hc/e$.
The diagonal approximation for $\zeta^{(C)}$ then gives
\begin{equation}
\label{zdiag}
\zeta^{(C)} ({\bf  r}_1, {\bf  r}_2 ; \omega ;H)
= \sum_{t:{{\bf  r}_1} \rightarrow {{\bf r}_2}}
\tilde\zeta_t^{(C)} ({\bf  r}_1, {\bf  r}_2 ; \omega ;H) \; ,
\end{equation}
where
\begin{eqnarray}
&&\!\!\! \tilde\zeta_t^{(C)} ({\bf  r}_1 , {\bf  r}_2 ; \omega ;H)
\simeq
\frac{v_F |D_t|^2}{2\pi\bar{\nu} \ell} \times  \nonumber \\
&& \times 
\exp \left[- \frac{L_t}{L_\phi} - \frac{L_t}{\ell}  + i\omega T_t
+ i \frac{4\pi}{\varphi_0}
\int_{{\bf  r}_1}^{{\bf  r}_2} {\bf A.dr} \right] \; .
\label{zsc}
\end{eqnarray}
The level broadening (implicit in $L_\phi$, Eq.~(\ref{Lphi})) 
was introduced via $\omega \rightarrow \omega + i\gamma$. 
Eq.~(\ref{zsc}) depends, besides $\ell$, only on the system
without disorder and holds for both integrable and chaotic geometries.

\subsection{Disorder induced magnetic response}
\label{subsec:mag}

Using Eqs.~(\ref{def_sus}) and (\ref{Kdiag}) the disorder induced 
contribution to the average magnetic susceptibility is given by
($\varphi = H L^2/ \varphi_0$)
\begin{equation}
\frac{\left< \chi^d (\varphi) \right>}{\left| \chi_L \right|} \approx
- \frac{6}{\pi^2}  \frac{\partial^2}{\partial \varphi^2} \
\sum_{n=1}^{\infty} \frac{1}{n} {\cal S}_{n}^{(C)} (0 ;\varphi) \; ,
\label{stos}
\end{equation}
where the bulk Landau susceptibility is 
$\chi_L = -e^2/24\pi mc^2$ for spinless electrons.

Assuming that $\ell$ remains fixed as $k_FL$ changes, 
one sees from Eqs.~(\ref{zsc}) and (\ref{stos}) that 
$\left< \chi^d (\varphi) \right>$ contains the dimensionless 
variables $\gamma /\Delta$ and $k_FL$ only in the combination 
$(\gamma /\Delta ) /k_FL$, since they are absorbed into
the dimensionless variable $L_{\phi}/L$, Eq.~(\ref{Lphi}).
 It is now assumed that the ${\cal S}_n^{(C)}$ 
in Eq.~(\ref{stos}) contain diagonal terms only.
 Therefore the propagator $\zeta^{(C)}$ (Eq.~(\ref{zdiag})) is made up 
of a summation over all diagonal pairs of paths including boundary 
scattering between any two given impurities situated at 
${\bf  r}_1$ and  ${\bf  r}_2$.
 On taking the trace over $n$ propagators $\zeta^{(C)}$, one sees
that the field sensitive part of $S_n$, Eq.~(\ref{sn}), 
consists of a summation over flux-enclosing pairs of closed paths 
(in position space) involving $n$ impurities and an arbitrary number of 
boundary scattering events.
 An example of a pair of paths contributing to ${\cal S}_{4}^{(C)}$ 
is shown in Fig.~\ref{DIAG}b.

%
%
%
%
%
\section{Integrable geometries:\\ square billiards}
\label{sec:square}
In the following we apply the above formalism to compute the
magnetic response of an ensemble of disordered billiards with 
regular geometry. We illustrate the method and present numerical
results for the case of square billiards.

\subsection{Numerical technique}

We consider specular reflection at the boundaries and employ 
the extended zone scheme \cite{Gef:94,Ull:95} to write 
$\zeta^{(C)}({\bf  r}_1 , {\bf  r}_2 ; \omega ;H)$, (Eq.~(\ref{zdiag})),
as a sum of propagators along {\it straight line paths} 
$\tilde\zeta_t^{(C)} ({\bf  r}_1 , {\bf  r}_2^t ; \omega ;H)$, 
of the form Eq.~(\ref{zsc}), where ${\bf  r}_2^t$ are images of
the position ${\bf  r}_2$.
The diagrams ${\cal S}_n^{(C)}$ (Eq.~(\ref{sn})) 
are then calculated by diagonalising $\zeta^{(C)}$.

 At zero magnetic field this approach enables 
one to recover the result of Ref.~\cite{A+G:93} for the approximate 
diagonalisation of the spectral correlation function
$K^d(\varepsilon_1,\varepsilon_2 ;H=0)$ as outlined in the Appendix.

For finite magnetic field the integrals over the magnetic vector 
potential along the paths do not allow for an analytical 
diagonalisation of $\zeta^{(C)}$.
However we are able to use the fact that all the 
variations of $\zeta^{(C)}$ occur on classical lengthscales; rapid 
oscillations on the scale of $\lambda_F$ cancel out. This
enables an efficient numerical computation. To this end we
discretise the configuration space of the square billiard
using a lattice with grid size greater than $\lambda_F$.
By summing over all trajectories (up to a length $\gg L_\phi$) 
which connect two lattice cells  we compute the corresponding
matrix elements of $\zeta^{(C)}$ in this representation.
After diagonalization we obtain $\left< \chi^d (H) \right>$ 
from Eq.~(\ref{stos}) \cite{note4}. This method is not restricted
to the square geometry but can be in principle applied to any geometry.
%
%
%
\begin{figure}
\vspace{0.3cm}
\hspace{0.02\hsize}
\epsfxsize=0.9\hsize
\epsffile{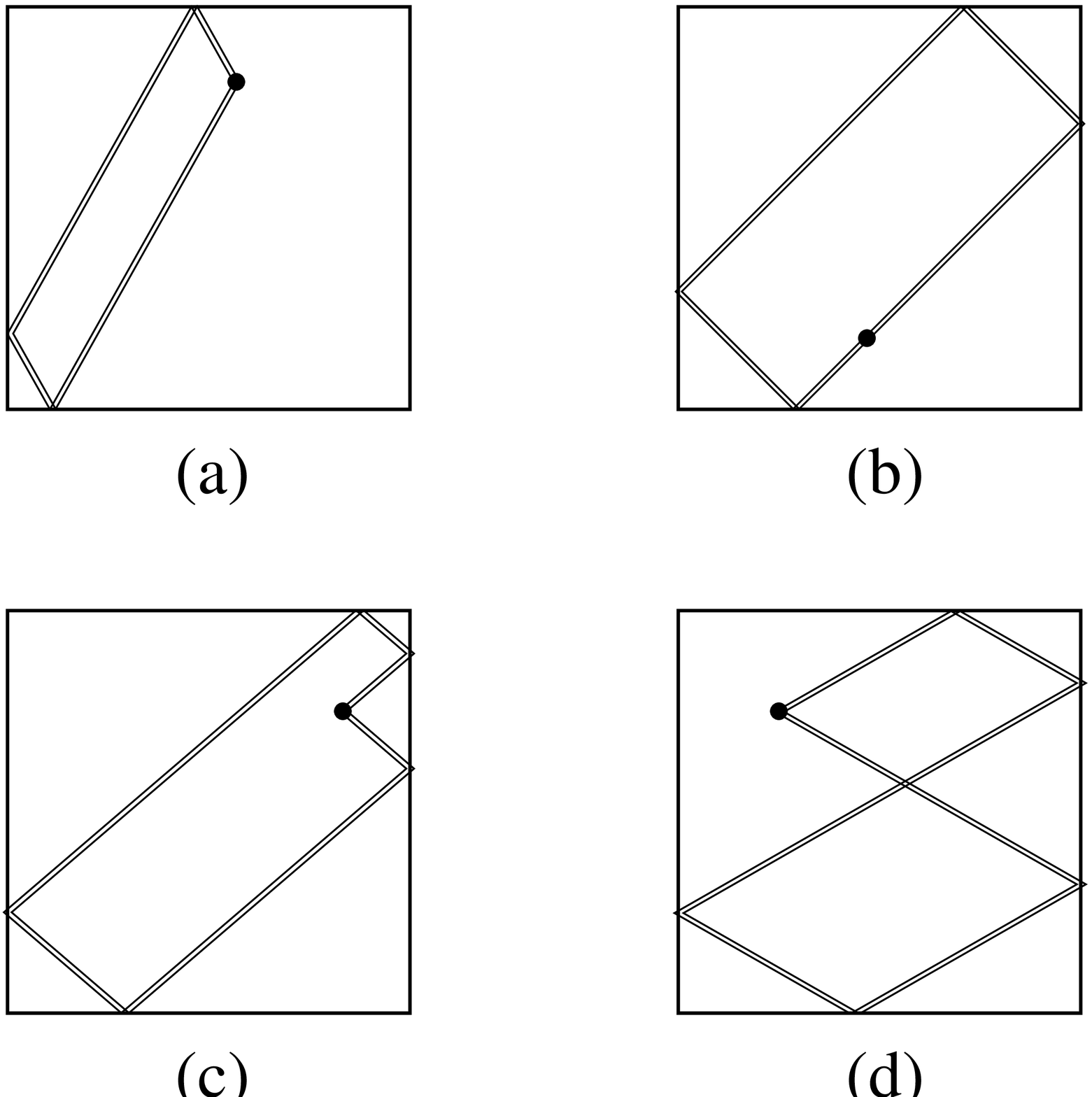}
\vspace{0.3cm}
 \refstepcounter{figure}
\label{ORBITS}
{\small \setlength{\baselineskip}{10pt} FIG.\ \ref{ORBITS}.
Examples of diagonal pairs of real-space trajectories with one 
impurity scattering position in a square.
 (a), (c), (d) show pairs of typical recurrent orbits 
with momentum exchange at 
the impurity which contribute to $\left<\chi^d (H)\right>$,
 whereas (b) shows a pair of orbits following periodic orbits of the 
corresponding clean system which contribute to $\left<\chi^L (H)\right>$.}
\end{figure}
%
%
Note that the diagrams ${\cal S}_n^{(C)}$ do not include closed 
pairs of paths which follow periodic orbits of the corresponding 
clean system.
 Such paths involve zero momentum transfer between 
the Green functions at the impurity positions. They actually represent 
disconnected diagrams and are included in $\left<\chi^L (H)\right>$
which has been computed in Ref.~\cite{Ric:96}.
 It is the presence of these periodic orbits in the determination of
$\left<\chi^L (H)\right>$ that leads to strong sensitivity of the
orbital magnetism with respect to 
the system geometry \cite{Ri:96b,Ull:95,note5}.
 The above numerical procedure does not distinguish between 
connected contributions to ${{\cal S}_1}$, with momentum exchange 
at the impurity as in Figs.~\ref{ORBITS}a,c,d, and disconnected diagrams 
involving one common impurity position.
 The dominant disconnected diagrams arise from the 
shortest flux enclosing periodic orbits of length $L_t = 2 \sqrt{2}L$, 
Fig.~\ref{ORBITS}b, and their repetitions.
 An estimate of their magnitude is obtained following the approach of
RUJ \cite{Ric:96,Ull:95}.
 On summation over all repetitions of the fundamental orbit we find
\begin{equation}
\frac{\left< \chi (0) \right>_{\rm dis}}{\left| \chi_L \right|} 
\simeq
\frac{8\sqrt{2}}{5\pi}\frac{L}{\ell}
\frac{1}{\sinh^2[\sqrt{2}(L/\ell + L/L_\phi)]} \; .
\label{susb}
\end{equation}
To avoid double counting we explicitly subtract 
this estimation from the numerical determination 
of $\left<\chi^d (H)\right>$.

%
%
\subsection{Zero field susceptibility}
We briefly summarize our results for the magnetic 
susceptibility at zero field and refer the reader to Ref.~\cite{McC+Ri:98}
for further details.
The technique covers the whole range from the diffusive regime
to the clean limit and yields an averaged 
susceptibility which is always paramagnetic.
In the diffusive limit, $\ell <L$, 
the disorder induced susceptibility increases linearly
with $\ell$ in agreement with Ref.~\cite{Oh:91}.
In the ballistic regime, $\left<\chi^d (0)\right>$ 
exhibits a maximum as a function of $\ell /L$.
The occurence of the maximum may be related to the competition
between different effects of the impurity scattering on 
$\zeta^{(C)}$, Eq.~(\ref{zsc}): while the single particle
Green functions are exponentially damped with $\ell$, it enters
as $\ell^{-1}$ into the prefactor.
%
%
%
\begin{figure}
\vspace{0.3cm}
\hspace{0.02\hsize}
\epsfxsize=0.9\hsize
\epsffile{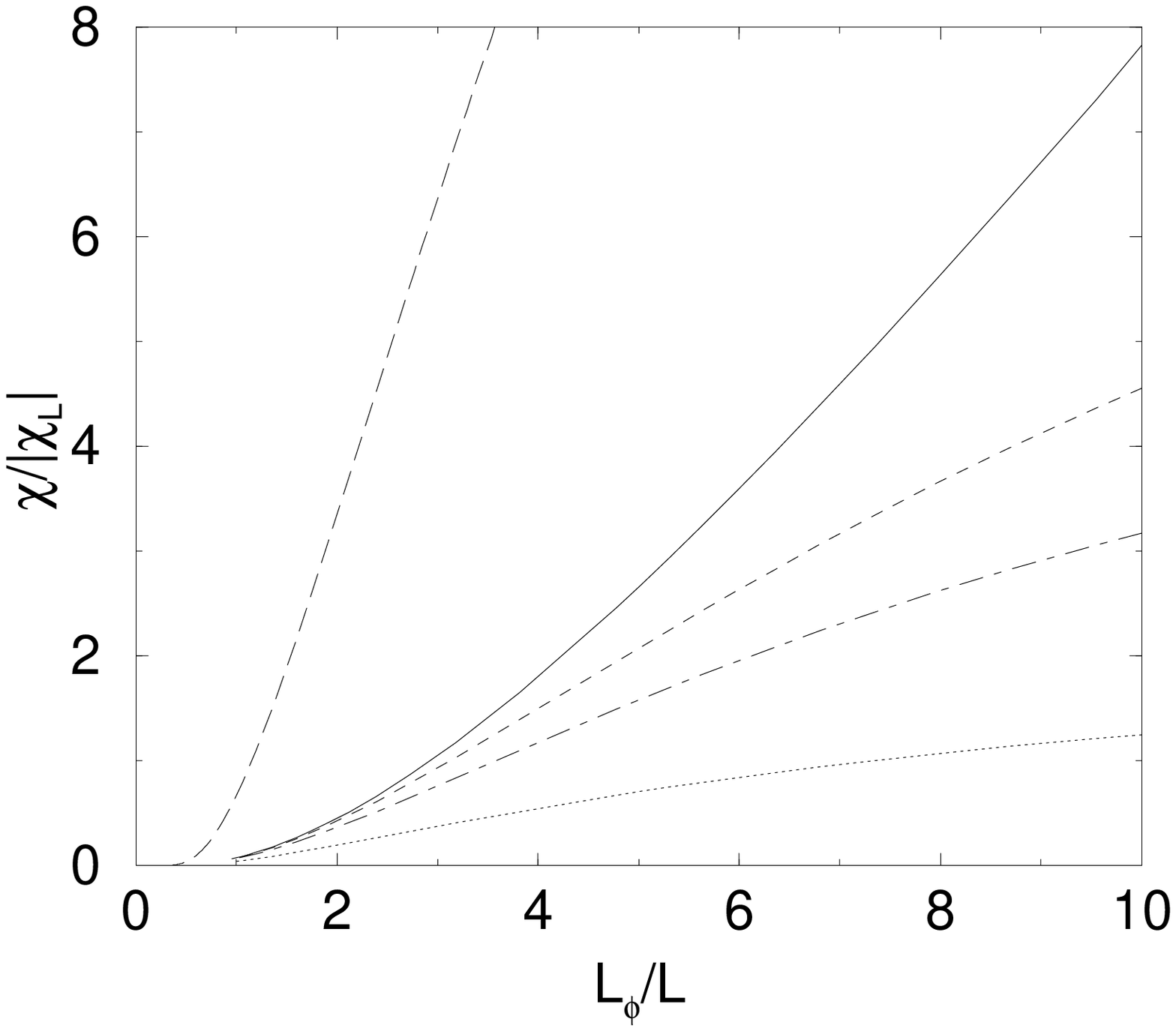}
\vspace{0.3cm}
 \refstepcounter{figure}
\label{GAM4}
{\small \setlength{\baselineskip}{10pt} FIG.\ \ref{GAM4}.
Comparison of the relative contributions of diagrams ${\cal S}_{n}^{(C)}$ 
to $\left<\chi^d (0)\right>$ for $\ell /L=4$ as 
a function of $L_{\phi} /L$.
 The solid line corresponds to the inclusion of all diagrams while, 
from the bottom, the dotted line corresponds to ${\cal S}_{1}^{(C)}$ only, 
the dashed-dotted line to ${\cal S}_{1}^{(C)}$ and ${\cal S}_{2}^{(C)}$, 
and the short dashed line to ${\cal S}_{1}^{(C)}$, ${\cal S}_{2}^{(C)}$ 
and ${\cal S}_{3}^{(C)}$.
 The long dashed line is the contribution of $\left<\chi^L (0)\right>$.}
\end{figure}
%
%
%
In the elastic regime, $L < \ell < L_\phi$,
$\left<\chi^d (0)\right>$ exhibits a weak dependence on $\ell$
but is on the whole close to the prediction by GBM
\cite{Gef:94}, which is a paramagnetic $\ell$-independent contribution
$ \left< \chi^d (0) \right>/\left| \chi_L \right| \approx 
0.23 \ k_FL (\Delta/\gamma) $,
For larger $\ell$, in the inelastic regime, $L, L_\phi < \ell$,
the susceptibility decays exponentially with both $\ell /L$ and 
$L/ L_\phi$.

The disorder induced susceptibility $\left<\chi^d (0)\right>$ for 
$\ell /L=4$ and a range of $L_{\phi} /L$ including both the elastic and 
inelastic regimes is shown as the solid line in Fig.~\ref{GAM4}.
We also include the contributions of diagrams ${\cal S}_{n}^{(C)}$,
Eq.~(\ref{sn}), with a different number $n$ of impurity scattering events
in order to analyse their relative weights.
From the bottom, the dotted line corresponds to ${\cal S}_{1}^{(C)}$ only, 
the dashed-dotted line to ${\cal S}_{1}^{(C)}$ and ${\cal S}_{2}^{(C)}$, 
and the short dashed line to ${\cal S}_{1}^{(C)}$, ${\cal S}_{2}^{(C)}$ 
and ${\cal S}_{3}^{(C)}$.
 For completeness, the long dashed line is the contribution of 
$\left<\chi^L (0)\right>$ (see below).
 The solid line curve shows linear behaviour in the elastic regime, 
as predicted by GBM, and exponential behaviour in the inelastic regime.
 In the limit $L_{\phi} \ll \ell$, where $\left<\chi^d (0)\right>$ is
very small, the lowest $n$ diagrams dominate but for $L_{\phi} > \ell$ 
a number of diagrams contribute significantly. We note that,
contrary to what is stated in Ref.~\cite{A+G:93},
the contribution from ${\cal S}_{1}^{(C)}$ is not particularly
small through the whole range of $L/L_\phi$.

%
%
\subsection{Finite field susceptibility}
\begin{figure}
\vspace{0.3cm}
\hspace{0.02\hsize}
\epsfxsize=0.9\hsize
\epsffile{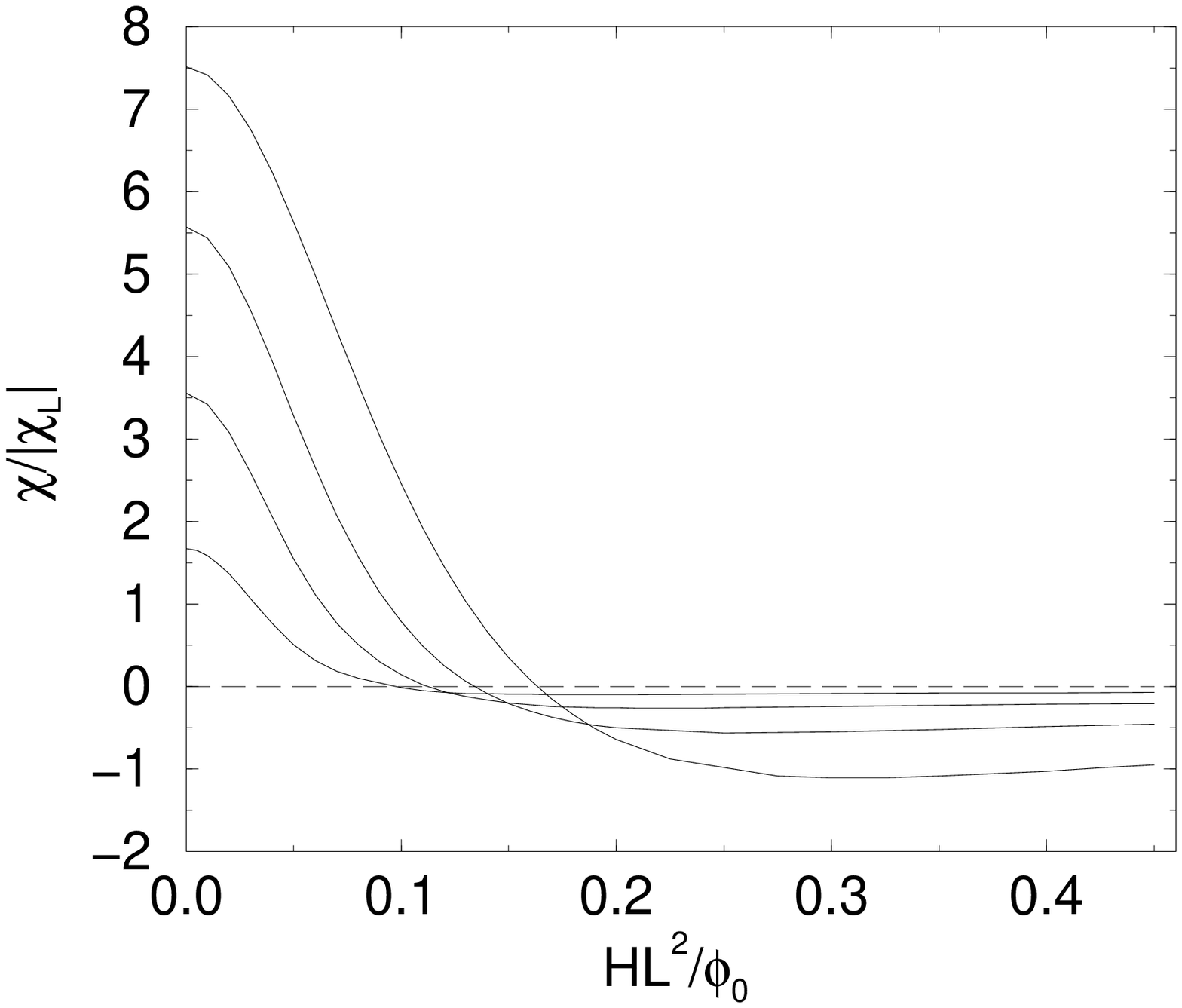}
\vspace{0.3cm}
 \refstepcounter{figure}
\label{HPLOT}
{\small \setlength{\baselineskip}{10pt} FIG.\ \ref{HPLOT}.
Field dependent susceptibility $\left<\chi^d (H)\right>$ 
for a square for $k_F L =60$ and 
$\gamma /\Delta = 1$ ($L_{\phi} /L = 9.55$).
 From the top (at $H=0$) the mean free path is 
$\ell /L=5, 10, 20$ and $50$.}
\end{figure}

We compute the finite field susceptibility $\left<\chi^d (H)\right>$ for 
weak fields, assuming that the field does not perturb trajectories 
away from their zero-field straight line paths, {\it i.e.\/} we 
assume $r_c > \ell$ where $r_c = \hbar k_F c/ eH$ is the cyclotron radius.
Fig.~\ref{HPLOT} shows $\left<\chi^d (H)\right>$ for $k_F L =60$ and 
$\gamma /\Delta = 1$ ($L_{\phi} /L = 9.55$).
 From the top (at $H=0$) the mean free path is 
$\ell /L=5, 10, 20$ and $50$.
 As the field increases from $H=0$ the susceptibility falls in 
magnitude and eventually becomes diamagnetic.
 The field value at which the susceptibility crosses over to diamagnetic 
behaviour decreases as $\ell /L$ increases.
 This implies that the typical area enclosed by 
a pair of recurrent orbits increases with $\ell$.

 In the range of validity of the above approach, $\ell < r_c, k_F L^2$ 
and $\gamma \agt \Delta$ we do not observe oscillations of 
$\left<\chi^d (H)\right>$ as a function of field as seen by 
GBM \cite{Gef:94} (albeit for a smaller value of $\gamma /\Delta$).
%
%
%
%
%
%
%
%
\subsection{Comparison of the combined semiclassical 
contributions with quantum mechanical results}
In this section we compare semiclassical results for
$\left<\chi^d (H)\right>$ and $\left<\chi^L (H)\right>$ with numerical 
quantum calculations taken from Ref.~\cite{Ric:96}.
In order to obtain an analytic expression for the contribution of 
size-induced correlations $\left<\chi^L (H)\right>$~\cite{McC+Ri:98},
we employ the results of RUJ \cite{Ric:96,Ull:95} 
at zero temperature and introduce 
the level broadening $\gamma$ in the same way as for the disorder 
correlations.
 It has been shown \cite{Ri:96b,Ull:95} that the low field
susceptibility of an ensemble of {\em clean} squares is dominated by the
shortest flux enclosing periodic orbits of length $L_t = 2 \sqrt{2}L$
and their repetitions over a broad range of temperature (or correspondingly
inelastic scattering strengths). For the ballistic white noise case
considered here, the effect of disorder averaging
 on the susceptibility was described by an additional damping
$\exp ( - L_t /\ell )$ of the response of the clean system \cite{Ric:96}.
This result corresponds to $\left<\chi^L (H)\right>$ including
the disorder damping $\exp ( - L_t /2\ell )$ of the two
single particle Green functions.
%
%
\begin{figure}
\vspace{0.3cm}
\hspace{-0.02\hsize}
\epsfxsize=0.9\hsize
\centerline{\epsffile{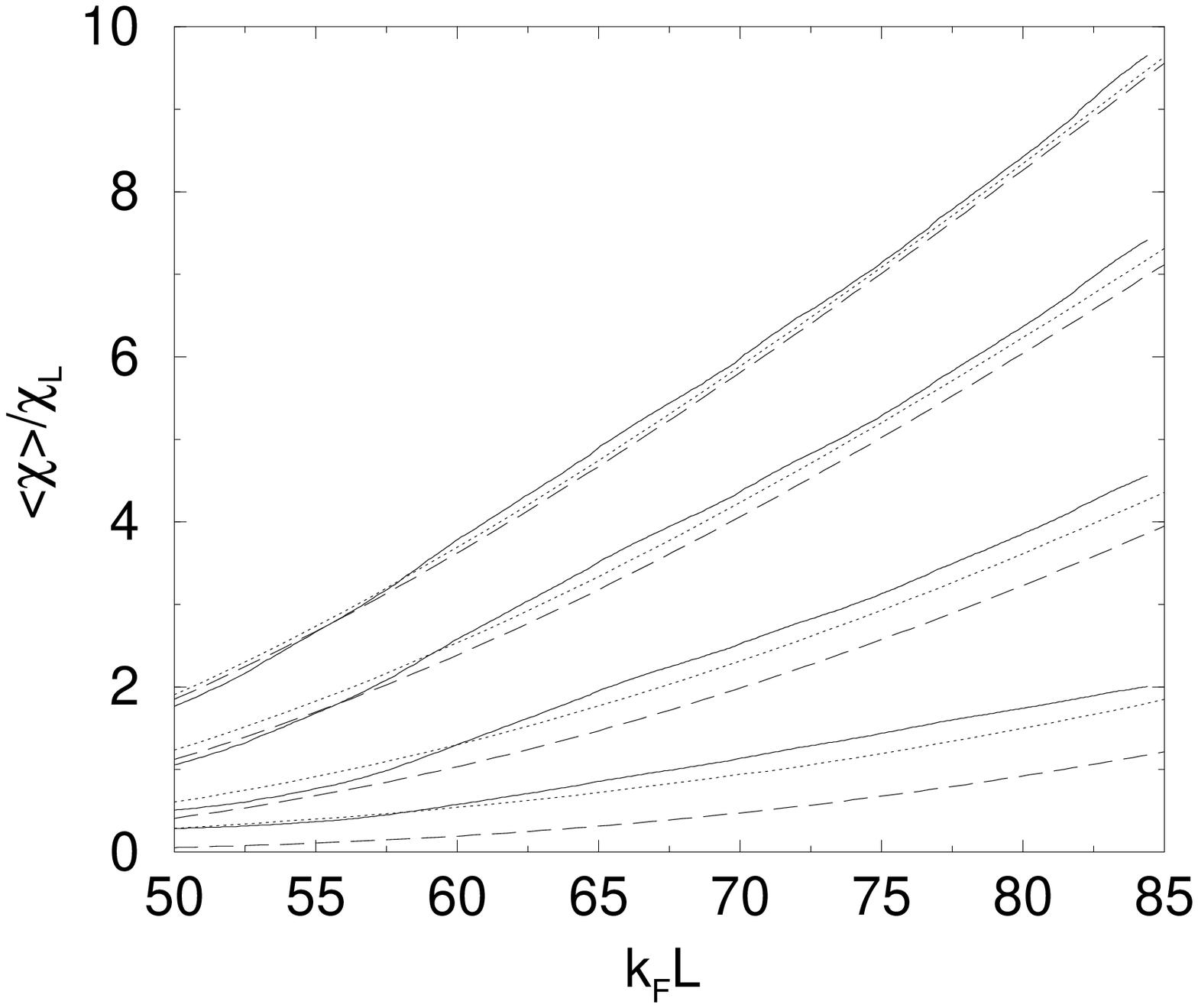}}
\vspace{0.2cm}
 \refstepcounter{figure}
\label{QUAN}
{\small \setlength{\baselineskip}{10pt} FIG.\ \ref{QUAN}.
Disorder- and size-induced contributions to the susceptibility.
The mean susceptibility for small field, $\varphi /\varphi_0 = 0.15$, 
and high temperature $k_BT /\Delta = 2$ ($L_T/L \approx 1.8$ 
at $k_FL=70$), for various disorder strengths, taking into 
account the variation of the mean free path with $k_F$; 
$\ell \propto k_F$.
 From the top, the elastic mean free path is $\ell /L = 8,4,2,1$ 
(at $k_FL \approx 70$).
The full and dashed curves are data from a quantum mechanical 
calculation and from a semiclassical evaluation of the contribution 
of size-induced correlations, respectively, taken from Ref.~\cite{Ric:96}.
 The dotted curve represents the sum of our semiclassical evaluation of the 
disorder-induced correlations, using 
$\gamma /\Delta \approx \pi k_BT /\Delta$, and
the semiclassical data for size-induced correlations 
from Ref.~\cite{Ric:96}.}
\end{figure}
%
%

The numerical quantum calculations~\cite{Ric:96} were obtained by 
diagonalisation of the Hamiltonian for non-interacting spinless 
particles in a square billiard with perpendicular magnetic field 
and white noise random potential.
 In this calculation temperature smearing, rather than level 
broadening due to inelastic effects, was introduced.

 The full lines in Fig.~\ref{QUAN} show the numerical results 
for various disorder strengths as a function of $k_FL$.
 From the top, the elastic mean free path is $\ell /L = 8,4,2,1$ 
(at $k_FL \approx 70$).
 As $k_FL$ changes, the mean free path is assumed to change also, 
according to $\ell \propto k_F$.

The dashed curves in Fig.~\ref{QUAN} are data taken from 
Ref.~\cite{Ric:96} for the semiclassical evaluation of the contribution 
of size-induced correlations to the susceptibility, 
$\left< \chi^L (H) \right>$.
 We calculated the contribution of the disorder-induced correlations, 
$\left< \chi^d (H) \right>$, 
using the approximation\cite{noteTemp}
$\gamma /\Delta \approx \pi k_BT /\Delta$ and 
taking into account the variation of mean free path with $k_FL$.
 The resulting semiclassical evaluation of the total susceptibility, 
$\left< \chi^L (H) \right> + \left< \chi^d (H) \right>$, 
is obtained by adding our results to those given by the dashed curves; 
the results are shown as dotted curves in Fig.~\ref{QUAN}.

For $k_FL \agt 60$ the agreement of the semiclassical (dotted) curves 
with the quantum results is considerable; it is clear that the 
difference between the semiclassical results for 
$\left< \chi^L (H) \right>$ and the quantum results is accounted for 
by the addition of $\left< \chi^d (H) \right>$.
 However even for $k_FL \agt 60$ the agreement is not exact; we 
believe that this minor discrepancy is due to the different mechanisms 
of damping that were used.

For $k_FL \alt 60$ there is worse agreement between the 
semiclassical (dotted) curves and the quantum results. This
may be related to the fact that the semiclassical results 
show no sign of the oscillatory structure present in the quantum curves.
This structure may be caused by off-diagonal pairs
of different families of
short periodic orbits contributing to $\left< \chi^L (H) \right>$,
which are not completely suppressed through energy average and
are not included in the present semiclassical approach.

\subsection{Relation to experiment }

Measurements of the orbital magnetism of ballistic systems, which
are experimentally realized as semiconductor microstructures, are still rare
\cite{Mailly,Lev:93}.
In the experiment of Ref.\cite{Lev:93} the susceptibility of ensembles of
squares was studied. The samples used
 had estimated values for the elastic mean free path
of $\ell /L \sim 1-2$, for the phase-coherence length
of $\sim (3-10) L $ and for the thermal cutoff length
of $L_T /L \sim 2$. Therefore, the lengthscale $L_\phi$ 
(Eq.~(\ref{Lphi})) is determined by the shorter length $L_T$. 

Fig.~\ref{QUAN} shows that for these experimental
parameters 
both the disorder and size-induced
correlations are relevant, however the latter contribution is dominant.

The above remarks hold for white noise disorder. However,
experimental ballistic structures as those of Ref.~\cite{Lev:93}
are usually characterised by smooth disorder potentials.
Smooth disorder effects can be in principle incorporated
into the present calculation by introducing an angle-dependent
cross section for the impurity scattering between two successive
trajectory segments.
The effect of smooth disorder on $\langle \chi^L \rangle$ has been analysed
in Ref.~\cite{Ric:96} which showed that the reduction of the
clean contribution is not as strong as that for white noise disorder
and no longer exponential. We therefore expect that 
in the parameter regime of the experiment the domination 
of the susceptibility by size-induced correlations is further
enhanced when considering smoothed disorder.
 
The measured value of the susceptibility at {\em low}
temperature was $\chi(0) \sim 100 |\chi_L|$,
with an uncertainty of about a factor of four.  The combined 
contributions $\langle\chi^d\rangle$ and $\langle\chi^L\rangle$,
together with an interaction contribution of the same order
\cite{Ull:97}, are in broad agreement with the experimental result.
We note, however, that a theoretical explanation of
the temperature dependence of the measured susceptibility
is still lacking.

%
%
%
\section{Chaotic geometries}
\label{sec:chaos}
For systems with a generic chaotic, clean counterpart we obtain
an analytical estimate for $\left<\chi^d (0)\right>$ under certain
statistical assumptions with respect to the classical trajectories involved.
To this end we make use of the relation\cite{AIS:93}
 ($L^2$ denotes the system area)
\begin{equation}
\label{arga}
\frac{\hbar}{\pi \bar{\nu}} \, \sum_{ j : {\bf r}_1 \to {\bf r}_2 }
|D_j |^2 \delta(t-t_j) = P({\bf r}_1,{\bf r}_2;t) \; 
\end{equation}
in order to transform the sums over classical densities $|D_t|^2$ 
in Eq.~(\ref{zsc}) into classical probabilities $P({\bf r}_1,{\bf r}_2;t)$
to propagate classically  between impurities at 
${\bf r}_1$  and ${\bf r}_2$ in time $t$. 

In the following we assume that in the ``ballistic'' 
regime $\ell ,L_{\phi} \gg L$  
the conditional probability $P({\bf  r}_1,{\bf  r}_2 ;t|{\cal A})$ to
accumulate  an ``area'' ${\cal A}$ during the propagation from
${\bf  r}_1$ and  ${\bf  r}_2$ 
is independent of ${\bf  r}_1$ and ${\bf  r}_2$.
Following Ref.~\cite{Ri:96b} which considered clean billiards, 
we find from Eqs.~(\ref{zsc}) and (\ref{arga})
\begin{eqnarray}
 \zeta^{(C)} ({\bf  r}_1 , {\bf  r}_2 ; \omega ;H)
&& \, \simeq
\frac{1}{\tau} \int_{0}^{\infty} dt 
\int_{-\infty}^{\infty} d{\cal A} \,
P({\bf  r}_1,{\bf  r}_2 ;t|{\cal A})\nonumber \\
&& \!\!\! \!\!\! \!\!\! \times 
\cos \left( \frac{4\pi{\cal A}H}{\varphi_0} \right)
\exp \left[- \gamma t - \frac{t}{\tau} + i\omega t  \right] \; .
\label{zsc2}
\end{eqnarray}
We assume that after several bounces off the boundary the area 
distribution $P({\bf  r}_1,{\bf  r}_2 ;t|{\cal A})$ becomes Gaussian 
with a variance $\sigma$ independent of ${\bf  r}_1$ and ${\bf  r}_2$ 
and independent of  $\ell$.
 Substituting the above expression for $\zeta^{(C)}$ into 
Eq.~(\ref{sn}) and performing the $r$, $t$ and ${\cal A}$ integrals 
we obtain the following closed approximate form for ${\cal S}_{n}^{(C)}$:
\begin{equation}
{\cal S}_{n}^{(C)}(\omega ;H) \approx \left\{
\frac{8\pi^2 H^2 \ell \sigma}{\varphi_{0}^2} + 1 + \gamma\tau - i\omega\tau
\right\}^{-n} \; .
\label{snch}
\end{equation}
Summation of the ${\cal S}_{n}^{(C)}$, Eq.~(\ref{stos}), leads to 
the disorder induced contribution to the average susceptibility,
\begin{equation}
\frac{\left< \chi^d (\varphi) \right>}{\left| \chi_L \right|} \approx
- \frac{24}{\varphi_{0}^2}  \frac{\partial^2}{\partial \varphi^2} \
\ln \left\{ 
\frac{\frac{8\pi^2 \varphi^2 \ell \sigma}{L^4} + \frac{\ell}{L_{\phi}}}
{1+\frac{8\pi^2 \varphi^2 \ell \sigma}{L^4} + \frac{\ell}{L_{\phi}}} 
\right\} \; ,
\label{stosch}
\end{equation}
where $\varphi = H L^2/ \varphi_0$.
 It is possible to obtain an explicit expression for 
$\left< \chi^d (\varphi) \right>$ analytically.
 For brevity we give only the limits.

The susceptibility has a paramagnetic maximum for zero field, 
\begin{equation}
\frac{\left< \chi^d (0) \right>}{\left| \chi_L \right|} 
\simeq \frac{96 \sigma L_{\phi}^2}{L^4(L_{\phi}+\ell )}  \; ,
\label{chdo}
\end{equation}
and it changes sign only once, for example in the elastic regime 
at the critical flux
\begin{equation}
\varphi_c^2 \approx \frac{L^3}{8 \pi^2 \sigma (L_{\phi}/L)} \; ,
\qquad   L_{\phi} > \ell \; .
\label{sign}
\end{equation}
For large field, $ \varphi \gg \varphi_c$,
 the susceptibility remains diamagnetic and decays rapidly,
\begin{equation}
\frac{\left< \chi^d (\varphi) \right>}{\left| \chi_L \right|} 
\sim -\frac{6 L^3}{\pi^4 \sigma  (\ell /L)  \, \varphi^4} 
\; .
\label{chinfin}
\end{equation}
%
%
\begin{figure}
\vspace{0.3cm}
\hspace{0.02\hsize}
\epsfxsize=0.9\hsize
\epsffile{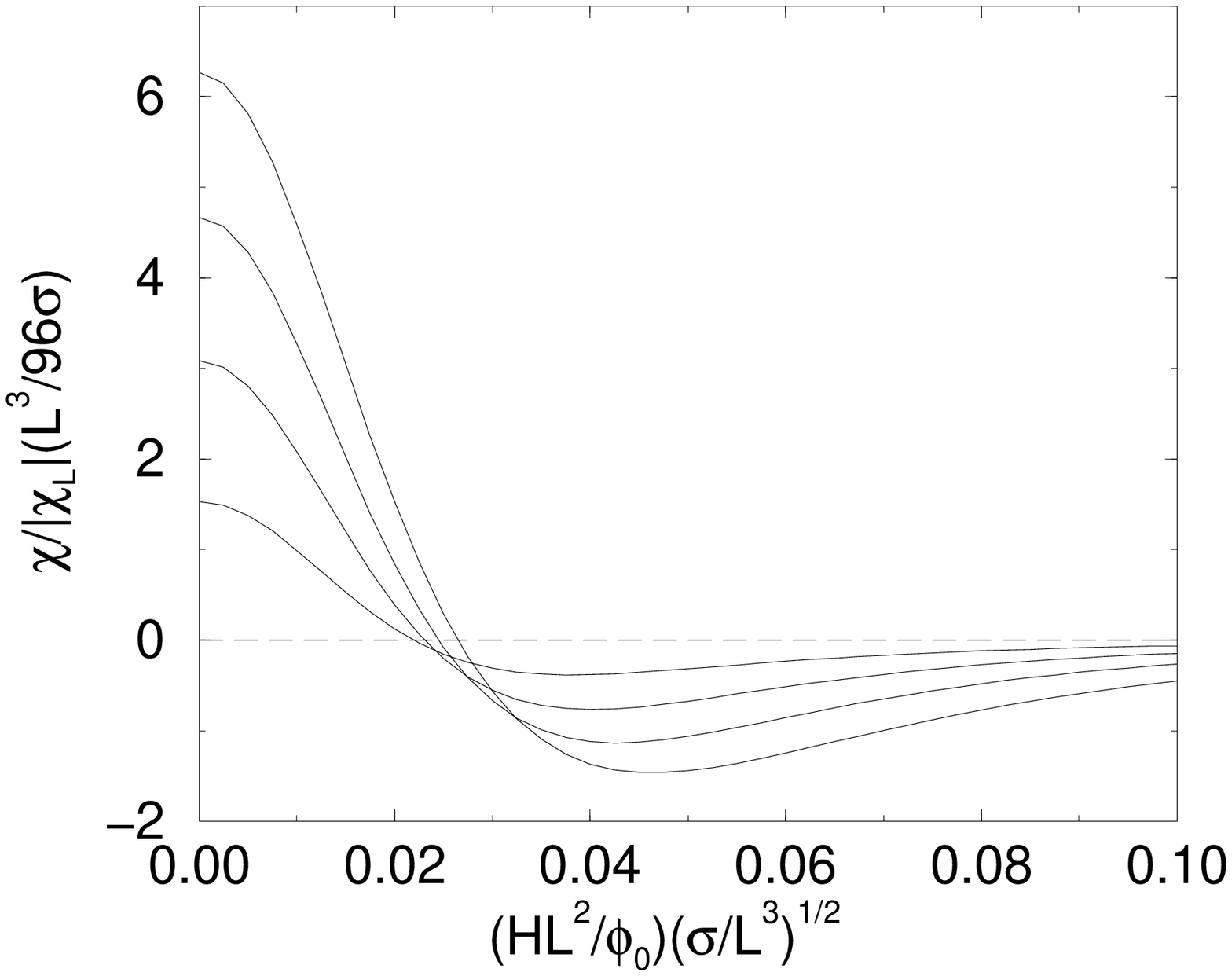}
\vspace{0.3cm}
 \refstepcounter{figure}
\label{CHPLOT}
{\small \setlength{\baselineskip}{10pt} FIG.\ \ref{CHPLOT}.
Field dependent susceptibility $\left<\chi^d (H)\right>$ 
(normalized) for a generic chaotic geometry for $k_F L =60$ and 
$\gamma /\Delta = 1$ ($L_{\phi} /L = 9.55$).
 From the top (at $H=0$) the mean free path is 
$\ell /L=5, 10, 20$ and $50$.}
\end{figure}
%
%
The susceptibility as a function of field is plotted in 
Fig.~\ref{CHPLOT} for $k_F L =60$ and 
$\gamma /\Delta = 1$ 
($L_{\phi} /L = 9.55$). 
 Various disorder strengths are shown, from the top (at $H=0$) 
$\ell /L=5, 10, 20$ and $50$.
These curves are qualitatively similar to those for 
the square geometry shown in Fig.~\ref{HPLOT}. This may be 
related to the fact that for trajectories being (multiply) scattered
at impurities the character of the clean geometry, namely regular
or chaotic, is of minor importance. Note however that the
classical variance $\sigma$ determines the height and the 
correlation field $\varphi_c$.
%
%
\begin{figure}
\vspace{0.3cm}
\hspace{0.02\hsize}
\epsfxsize=0.9\hsize
\epsffile{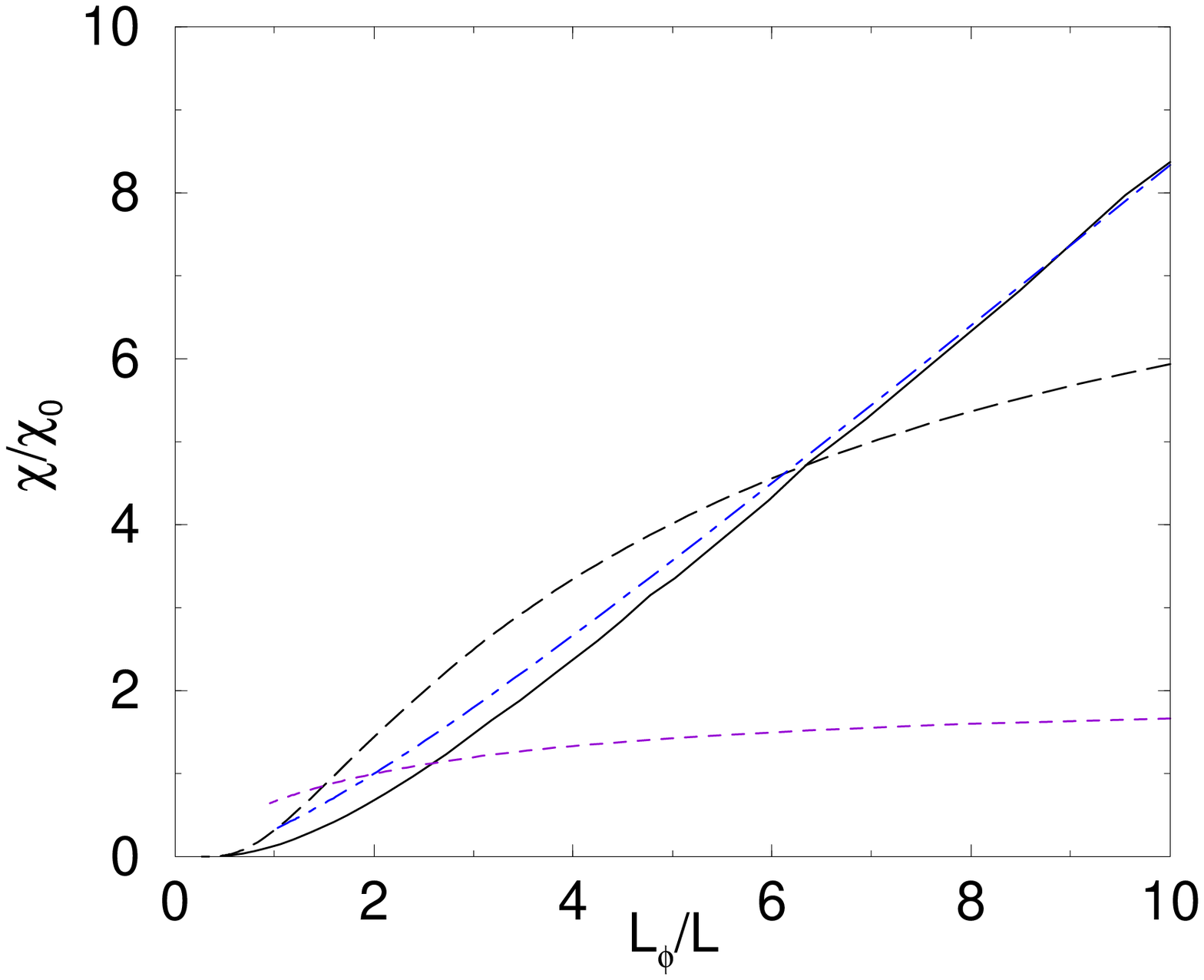}
\vspace{0.3cm}
 \refstepcounter{figure}
\label{GAMCHAOS2}
{\small \setlength{\baselineskip}{10pt} FIG.\ \ref{GAMCHAOS2}.
Contributions to the susceptibility for $k_F L = 60$ and $\ell/L = 2$,
with (normalized) 
semiclassical estimates for $\left< \chi^d (0) \right>$ 
(dashed-dotted line) and $\left< \chi^L (0) \right>$ (short dashed) 
for a generic chaotic geometry and for $\left< \chi^d (0) \right>$ 
(solid line) and $\left< \chi^L (0) \right>$ (long dashed) 
for square geometry. $\chi_0 = |\chi_{\rm L}|$ for the square and
$|\chi_{\rm L}|L^3/96\sigma$ for the chaotic geometry.}
\end{figure}
%
%
The susceptibility $\left<\chi^d (0)\right>$ as a function of $L_{\phi}$ 
for $\ell /L =2$ is shown as the dashed-dotted line in Fig.~\ref{GAMCHAOS2}
and a corresponding approximation for 
$\langle \chi^L(0) \rangle$ \cite{Ull:95} for finite $L_{\phi}$, 
\begin{equation}
\frac{\left< \chi^L (0) \right>}{\left| \chi_L \right|} 
\simeq \frac{96 \sigma L_{\phi} \ell}{L^4(L_{\phi}+\ell )}  \; ,
\label{chLo}
\end{equation}
is shown as the short dashed line in Fig.~\ref{GAMCHAOS2}.
The two contributions (\ref{chdo}, \ref{chLo})
add up to an $\ell$-{\em independent} magnetic response
\begin{equation}
\frac{\left<\chi(0)\right>}{|\chi_L|} \simeq \frac{96 \sigma L_\phi}{L^4}\; .
\end{equation}
which is the same as that for clean
chaotic systems, given the assumption of an $\ell$-independent variance.

For comparison, results for the square geometry as a
function of $L_{\phi}$ for $\ell /L =2$ are also
presented in Fig.~\ref{GAMCHAOS2}.
The solid line shows our numerical results for 
$\left<\chi^d (0)\right>$ and the long dashed line shows
$\left<\chi^L (0)\right>$ (see Ref.~\cite{McC+Ri:98} for more
details).
The $L_{\phi}$ dependence also shows qualitative similarity between
the disorder-induced susceptibility in the square and the chaotic
geometry.
Naturally, the square and the chaotic
billiard show quite different behaviour for the size-induced susceptibility.

For the square a crossover from domination by disorder-induced
correlations to domination by size-induced correlations
occurs for $L_{\phi} \sim \ell^2$, which contrasts with
chaotic geometries where the crossover occurs for $L_{\phi} = \ell$.

%
%
\section{Concluding remarks}
We considered disorder-induced spectral correlations of 
mesoscopic quantum systems in the non-diffusive regime. We
combined a diagrammatic treatment of the disorder with a
semiclassical approach for the Green functions involved.
This leads to the representation of the diagrams involved in terms of
(pairs of) classical paths in real space which undergo both (multiple)
scattering at the impurities and reflections at the system boundaries.
This hybrid-type approach enables us to perform disorder and energy averages
and to study the complete crossover from the diffusive to the
clean limit for arbitrary values of the elastic mean free
path $\ell$ and $L_{\phi}$ (smaller than $v_F t_H$). In 
particular it is applicable to ballistic micro-structures of, 
in principle, arbitrary geometry. The approach, presented here for
billiard systems, applies also to systems with potentials whose 
effect is then incorporated in the semiclassical Green functions.

We focussed on the effect of disorder-induced correlations
on the averaged magnetic susceptibility which is closely related to
the spectral number variance and which amounts to evaluate
Cooper type contributions to the correlator. 
Corresponding diffuson type diagrams
can be computed equivalently. As an example of a system with
integrable clean counterpart, we treated in detail the 
(experimentally relevant) square billiard. We showed that there are two 
distinct regimes of behaviour depending on the relative magnitudes of 
$\ell$ and $L_{\phi}$. Certain statistical assumptions on the
trajectories in ballistic systems with a generic chaotic geometry 
enabled us to derive approximate analytical expressions for
the disorder induced magnetic response. It turns out that the
magnetic field dependence and $L_\phi$-dependence of the 
susceptibility, at least qualitatively, resembles that of the
integrable case. This implies that the actual geometry of the clean system
is of minor importance for the disorder induced orbital magnetism
even in the ballistic regime.

%
%
%
\acknowledgments
We are grateful to Y.~Gefen, S.~Kettemann, D.~E.~Khmelnitskii, 
M.~Leadbeater, I.~V.~Lerner and P.~Walker for useful discussions.
We thank the Isaac Newton Institute for Mathematical Sciences, Cambridge,
where part of this research was performed.
\appendix
\section{Diagonalisation of $\zeta$ operators for zero field}

 In this appendix we show that within the semiclassical approach the 
analytical diagonalisation of the $\zeta$ operators, Eq.~(\ref{zc}), for a
square geometry at zero field
is possible and straightforward. We employ the
 extended zone scheme which is constructed by reflecting 
the original billiard with respect to its boundaries.
 Any given trajectory connecting ${\bf  r}_1 \equiv (\xo,\yo)$ 
and ${\bf  r}_2 \equiv (\xt,\yt)$ involving 
an arbitrary number of specular reflections at the boundaries is tranformed
into a straight line path connecting ${\bf  r}_1$ and ${\bf  r}_2^t$ where 
${\bf  r}_2^t \equiv (\pm \xt + 2n_xL,\pm \yt + 2n_yL)$ is an image of 
the position ${\bf  r}_2 \equiv (\xt,\yt)$, $n_x$ and $n_y$ are integers.
 As a result a semiclassical propagator, 
$\zeta^{(C)} (\xo ,\yo ;\xt ,\yt )$,  may be written as a sum 
of propagators, ${\tilde\zeta^{(C)}} (\xo ,\yo ;\xt^t ,\yt^t )$,
along straight-line trajectories,
                  \ifpreprintsty
                   \else 
			\end{multicols}\vspace*{-3.5ex}{\tiny
                   \noindent\begin{tabular}[t]{c|}
                   \parbox{0.493\hsize}{~} \\ \hline \end{tabular}}
                                     \fi
\begin{equation}
\zeta^{(C)} (\xo ,\yo ;\xt ,\yt ) 
= \sum_{n_x=-\infty}^{\infty} \sum_{n_y=-\infty}^{\infty}
{\tilde\zeta^{(C)}} (\xo ,\yo ;\pm \xt + 2n_xL ,\pm \yt + 2n_yL).
\label{sums}
\end{equation}
%
%
%
Since ${\tilde\zeta^{(C)}}$ is dependent only on the distances 
$X \equiv (\xt^t-\xo)$ and $Y \equiv (\yt^t-\yo)$ along a straight line 
it is possible to introduce a Fourier transform 
$\lambda (q_x , q_y)$ and, using the Poisson summation rule, we may write
\begin{equation}
\zeta^{(C)} (\xo ,\yo ;\xt ,\yt ) 
= \sum_{m_x,m_y=-\infty}^{\infty}
 \int_{-\infty}^{+\infty}
{dq_x dq_y\over(2\pi)^2}\, 4\, \lambda (q_x , q_y)
\cos(q_x\xo)\cos(q_y\yo) 
\delta \left( \frac{q_xL}{\pi}\!-\! m_x \right) 
\delta \left( \frac{q_yL}{\pi}\!-\! m_y \right) 
e^{\left\{ iq_x\xt + iq_y\yt \right\}} . 
\label{sums2}
\end{equation}
Performing the integrals gives quantised momentum values 
$q_x=m_x\pi /L$, $q_y=m_y\pi /L$ and
\begin{eqnarray}
\zeta^{(C)} (\xo ,\yo ;\xt ,\yt ) &=& \frac{1}{L^2}\left[ \lambda (0,0) + 
4 \sum_{m_x=1}^{\infty} 
\sum_{m_y=1}^{\infty}\lambda (\frac{m_x\pi}{L},\frac{m_y\pi}{L}) 
\cos(\frac{m_x\pi\xo}{L})\cos(\frac{m_y\pi\yo}{L})
\cos(\frac{m_x\pi\xt}{L})\cos(\frac{m_y\pi\yt}{L})\right. \nonumber \\
& + & \left. 2\sum_{m_x=1}^{\infty} 
 \lambda (\frac{m_x\pi}{L},0) \cos(\frac{m_x\pi\xo}{L}) 
      \cos(\frac{m_x\pi\xt}{L})
+ 2\sum_{m_y=1}^{\infty} 
 \lambda (0,\frac{m_y\pi}{L}) 
\cos(\frac{m_y\pi\yo}{L})\cos(\frac{m_y\pi\yt}{L}) \right] \; .
\label{sums4}
\end{eqnarray}
%
%
%
%
                   \ifpreprintsty
                   \else
		 {\tiny\hspace*{\fill}\begin{tabular}[t]{|c}\hline
                    \parbox{0.49\hsize}{~} \\
                   \end{tabular}}\vspace*{-2.5ex}\begin{multicols}{2}\noindent
                    \fi
Although $\zeta^{(C)} (\xo ,\yo ;\xt ,\yt )$ is not translationally 
invariant, it does have a periodicity of $2L$.
 The diagrams ${\cal S}_{n}$ may be written in 
terms of the Fourier transform $\lambda(q_x,q_y)$ by inserting 
the above expression into Eq.~(\ref{sn}).
 Spatial integrals are performed using the following relation,
\begin{equation}
\int_{0}^{L} \! \frac{dx}{L}
\cos(\frac{m_{x}\pi x}{L})\cos(\frac{m_{x^{\prime}}\pi x}{L})
= \frac{1}{2}\! \left( \delta_{m_{x},m_{x^{\prime}}} \!+\!
\delta_{m_{x},-m_{x^{\prime}}} \right),
\label{cosint}
\end{equation}
and the result is
\begin{equation}
{\cal S}_{n}^{(C,D)} = 
\sum_{m_{x}=0}^{\infty} \sum_{m_{y}=0}^{\infty}
\left[ \lambda (\frac{m_{x}\pi}{L} ,\frac{m_{y}\pi}{L}) \right]^n.
\label{dia}
\end{equation}
In order to evaluate $\lambda (q_x , q_y)$ we
Fourier transform the expression for the semiclassical operator 
$\tilde\zeta_t^{(C)}$, Eq.~(\ref{zsc}),
performing integrals over $X$ and $Y$.
 By changing from the cartesian coordinates $(X,Y)$ to polar coordinates 
$(r,\theta )$, the radial integration is done giving
\begin{eqnarray}
\label{thint}
\lefteqn{
\lambda (q_x , q_y) = } \\
& &  \frac{1}{2\pi} \int_{0}^{2\pi}
\frac{d \theta}{\left( 1+\gamma\tau -i\omega\tau\right)
+ i\ell \left( q_x \cos\theta + q_y\sin\theta \right)}, \nonumber
\end{eqnarray}
and we find
\begin{equation}
\lambda (q_x , q_y) = 
\frac{1}{\sqrt{\left( 1+\gamma\tau -i\omega\tau\right)^2 
+ \left( \ell q \right)^2}},
\label{zfinal}
\end{equation}
where $q^2 = q_x^2 + q_y^2$. This result is in agreement with 
the approximate diagonalisation of Ref.~\cite{A+G:93}. Together
with Eq.~(\ref{Kdiag}) it gives the spectral correlation function
in the ``ballistic'' regime.
%
%
%
%
%
%
%
%
%
%

%
%

\end{multicols}

\begin{references}
\bibitem{note3}
In a ballistic system such additional
oscillatory terms will however remain upon pure disorder average 
for fixed size \cite{A+G:93,Ric:96}.
\bibitem{note4}
A similar calculation has been applied to a different
``classical'' operator, describing interaction effects
in ballistic quantum dots, in Ref.~\cite{Ull:97}.
\bibitem{noteTemp} 
We note that a proper consideration\cite{M+R} of temperature smoothing
by convoluting  the spectral correlation functions with
the Fermi function amounts to replace the exponential damping
with $L_\phi$ in Eq.~(\ref{S_expan}) by the damping $x/\sinh x$ with
$x = \pi T_t k_BT/\hbar$ where $T_t$ is the period of the classical
path involved and $T$ the temperature.
\end{references}
\end{document}